\documentclass[12pt]{iopart}
\expandafter\let\csname equation*\endcsname\relax
 \expandafter\let\csname endequation*\endcsname\relax  
\usepackage{amsmath}
\usepackage{amssymb}

\begin{document}

\title[]{On the late phase of relaxation of two-dimensional fluids:
turbulence of unitons}

\author{F. Spineanu and M. Vlad}

\address{National Institute of Laser, Plasma and Radiation Physics \\
Magurele, Bucharest 077125, Romania}
\ead{florin.spineanu@inflpr.ro}
\vspace{10pt}
\begin{indented}
\item[]October 2016
\end{indented}

\begin{abstract}
The two-dimensional ideal fluid and the plasma confined by a strong magnetic
field exhibit an intrinsic tendency to organization due to the inverse
spectral cascade. In the asymptotic states reached at relaxation the
turbulence has vanished and there are only coherent vortical structures. We
are interested in the regime that precedes these ordered flow patterns, in
which there still is turbulence and imperfect but robust structures have
emerged. To develop an analytical description we propose to start from the
stationary coherent states and (in the direction opposite to relaxation)
explore the space of configurations before the extremum of the functional
that defines the structures has been reached. We find necessary to assemble
different but related models: point-like vortices, its field theoretical
formulation as interacting matter and gauge fields, chiral model and
surfaces with constant mean curvature. These models are connected by the
similar ability to describe randomly interacting coherent structures. They
derive exactly the same equation for the asymptotic state (sinh-Poisson
equation, confirmed by numerical calculation of fluid flows). The chiral
model, to which one can arrive from self-duality equation of the field
theoretical model for fluid and from constant mean curvature surface
equations, appears to be the suitable analytical framework. Its solutions,
the unitons, aquire dynamics when the system is not at the extremum of the
action. In the present work we provide arguments that the underlying common
nature of these models can be used to develop an approach to fluid and
plasma states of turbulence interacting with structures.
\end{abstract}

%

%
%
%

\section{Introduction}

\bigskip

A characteristic of the dynamics of the plasma immersed in magnetic field is
the dominant role of two space lengths: of the gradient of parameters
(density, temperature), $L$, and of the Larmor gyration $\rho _{L}$. The
first governs the reservoir of free energy and the second the rate of
extraction of this energy through instabilities. A strong magnetic field
determines quasi-two-dimensionality which favors the inverse spectral
transfer, \emph{i.e.} of energy from small to large spatial scales. This is
manifested by a tendency to generate coherent pattern of flow, at the scale $%
L$ (convective cells in confined plasma). Then, although the first
manifestation of the nonlinearities is the turbulence, there are also
robust, quasi-coherent structures of flow. The mixture of turbulence and
structures depends on the magnitude of the drive and of the region of
spectrum where energy is injected. It is only at relaxation (absence of
sources) from an initial turbulent state that the ordered pattern of flow
produced by the inverse spectral cascade is clearly visible. This evolution
is very complex and it may be useful to benefit from the study of the
similar problem for the ideal $2D$ neutral (Euler) fluids. The absence of $%
\rho _{L}$ rises doubts on the possibility to draw such a parallel, as the
complex set of equations (fluid and kinetic) of the plasma are mapped to the
non-dissipative Navier-Stokes (\emph{i.e.} Euler) equation. There are
however many cases where the Euler equation appears to be an adequate
description of the convective processes in plasma \cite{hassamdrake1}, \cite%
{guzdarshear}.

quasi-coherent structures are a component of the fluctuating field even deep
in the turbulent regime. In general they are randomly and transiently
generated, being a signature of a tendency of organization that competes
with the turbulence. Experiments and numerical simulations show that for
ideal $2D$ Euler fluid the relaxation from an initial turbulent state leads
to a highly coherent structure of flow. The process consists of separation
and clusterization of the positive respectively negative vorticity via
like-sign vortex mergings. The asymptotic state consists of a dipole and it
is known that the streamfunction verifies the sinh-Poisson equation.

\bigskip

The late phase of the relaxation, \emph{i.e.} the regime that precedes the
final ordered state, is a mixture of coherent structures and turbulence.
This state cannot be studied with methods of statistical physics since the
number of active degrees of freedom (as shown for example by a
Karhunen-Loeve analysis) decreases while the number of cumulants
(irreducible correlations) increases, due to progressive formation of
quasi-coherent structures. Then any perturbative expansion intended to
remain close to the Gaussian statistics cannot converge \cite{krommes20021}.
The study of regimes where turbulence and robust coherent structures are
simultaneously present requires a framework in which they both can be
represented analytically. Since they are as different as random fluctuations
versus solitons/vortices, the specific analytical methods (statistical
physics versus geometric-algebraic methods of exact integrability) should
find a common platform. This is possible since, individually, the turbulent
field and the coherent structures are extrema of an unique action
functional. This can be written starting from the basic equations and
applying procedures like Martin-Siggia-Rose method \cite{martinsiggiarose}.
It is more transparent to use path integral formulation \cite{jensenpath}
and to define a generating functional from which the correlations (turbulent
field in the presence of structures) can be obtained by functional
derivation. This has been applied for a drift wave vortex perturbed by
turbulence \cite{florinmadi2000} and for a collection of quasi-coherent
vortices embedded in a weak drift wave turbulent field \cite%
{florinmadistatist2005} in magnetized plasma. Technically, this approach
benefits of the powerful Feynman diagrammatic methods but the analytical
developments can become cumbersome and the applications are limited.

In this work we examine the advanced phases of relaxation of the
two-dimensional ideal (Euler) fluid. This state consists of few robust
vortices interacting with the turbulent field \cite{Montgomery1991}, \cite%
{Montgomery1992}. In the asymptotic state the turbulent field has vanishing
amplitude and only the static large scale vortical dipolar structure exist.
The streamfunction verifies the sinh-Poisson equation. In the following we
adopt this as the reference state and try to explore the regime that
precedes it: there is still turbulence, quasi-coherent structures have
emerged and the streamfunction does not yet verify the sinh-Poisson equation.

The reference state suggests for the $2D$ Euler fluid connections that may
prove to be useful: (1) statistical physics of point-like vortices (SPV),
(2) non-Abelian field theory (FT) of matter interacting with a Chern-Simons
gauge potential, and (3) surfaces with constant mean curvature (CMC) in the $%
3D$ Euclidean space. For each system the function that describes the state
verifies the sinh-Poisson equation and the corresponding (asymptotic) states
are as exceptional as our reference state for fluid. The systems evolve to
the highly organized asymptotic states driven by different mechanisms:
inverse cascade for the fluid; extremum of entropy at negative temperature
for the statistical ensemble of point-like vortices; extremum of the action
functional for the field theory model; curvature flow with area dissipation
at fixed volume, \emph{i.e.} minimization of the capillarity energy for
surfaces. Regarding the asymptotic states, their special nature has
individual characterization for these systems: static dipolar coherent
structure of flow, for the $2D$ fluid (the reference state); maximum
combinatorial entropy at fixed energy, for point-like vortices; self-duality
state, for the FT model; Constant Mean Curvature, for the surface. As
different as they may seem, the dynamical evolution toward the final states
and respectively the characteristics of the final states are however
strongly related. The mappings that can be established between them give us
the hope that asking a question in one framework we can get a hint from
another one, if it happens that the formulation is more favorable there.

To examine in parallel the asymptotic states (dipolar vortex, Self-Duality,
Constant Mean Curvature) based on their similar nature can be a useful
instrument of investigation and an interesting problem in itself. However
our objective is to explore the regime that precedes the limiting
(\textquotedblleft final\textquotedblright , \emph{i.e.} asymptotic) states,
when turbulence is still active. Or, the final states are all determined
from variational calculus which identifies the state but is unable to
describe the regime preceding it. Here however we are helped by the special
nature of the final states: like solitons or topological solutions, they are
robust and can be found in the preceding evolution regime even if they are
not exact solutions. There are interaction between them and with the
turbulent field. Models based on such inferred\ representation of special
solutions in regimes where they are just emerging are rather usual:
turbulence of Langmuir solitons in plasma \cite{langmuirturb}, turbulence of
topological defects in superconductivity \cite{shahmanton}, filaments of
nonlinear Schrodinger equation in optical turbulence \cite{mlejnekturbulence}%
, instanton plasma \cite{instantonplasma} or even random topological changes
in models of baryogenesis \cite{baryofluct}, \emph{i.e.} turbulence of
sphalerons. In our case the adequate framework would be the field
theoretical model with solutions of the self-dual state. Or, here there is a
problem. The self-dual states represent the absolute extremum of the action
functional of the FT model. Analytically they are described as solutions of
the sinh-Poisson equation, the same equation for the streamfunction of the
physical $2D$ Euler fluid. The equation is exactly integrable and solutions
are known. We can imagine using these solutions in the preceding regime as
coherent vortices interacting with the turbulent vorticity field. However
the exact state identified by the extremum of the action is self-duality and
the sinh-Poisson equation is derived under a particular hypothesis
(\textquotedblleft ansatz\textquotedblright ) about the algebraic content of
the matter and gauge fields. Then one can ask if, leaving the final
(reference) state and pulling-back to explore the regime which precedes it,
we are - or we are not - allowed to maintain this particular algebraic
structure. The functions that are not yet at exact self-duality may have an
algebraic content which is different of that simplified \textquotedblleft
ansatz\textquotedblright\ which has led us to the sinh-Poisson equation. On
the other hand we know that the asymptotic state is correctly identified
since it is confirmed that the physical fluid reaches the static structure
that is solution of this equation. The FT system, evolving toward SD will
have two tasks: to progressively alter the larger algebraic content by
reducing it to the simplified \textquotedblleft ansatz\textquotedblright\
and in the same time to decrease the magnitude of the action such that the
asymptotic state is static self-duality. If this is the case then we have to
consider not the exact solutions of the sinh-Poisson equation (since it has
been derived under a simplified \textquotedblleft ansatz\textquotedblright )
but the solutions of the self-duality equations that are less restrictive
with the algebraic content. These solutions exist and are called \emph{%
unitons}. Therefore we are led to consider the turbulence of the structures
manifested at self-duality, without adopting the simplified algebraic
\textquotedblleft ansatz\textquotedblright , and this means turbulence of
unitons.

We briefly review the models and show how it is derived the equation \emph{%
sinh}-Poisson in each of them. This will make plausible the mappings between
models. The connections between integrable equations, geometry of surfaces
and field theory have been much investigated in the mathematics and physics
literature. We limit ourselves to reveal some physical interpretations of
these connections which are potentially useful to the subject of turbulence
and structures in fluid flow. The purpose is to evaluate the possibility to
construct an effective analytical apparatus for the study of structures
embedded in turbulence.

\bigskip

\section{Euler fluid and system of point-like vortices}

The ideal (non-dissipative) incompressible fluid in two - dimensions, which
we will shortly call $2D$ Euler fluid, can be described by three functions $%
\left( \psi ,\mathbf{v,\omega }\right) $. From the scalar field
streamfunction $\psi \left( x,y,t\right) $ one derives the velocity vector
field $\mathbf{v}\left( x,y,t\right) =-\mathbf{\nabla }\psi \times \widehat{%
\mathbf{e}}_{z}$ (here $\widehat{\mathbf{e}}_{z}$ is the versor
perpendicular on the plane $\left( x,y\right) $) and the vorticity $\omega 
\widehat{\mathbf{e}}_{z}=\mathbf{\nabla \times v=}\Delta \psi \widehat{%
\mathbf{e}}_{z}$ . The Euler equation is the advection of the vorticity by
its own velocity field%
\begin{equation}
\frac{d\omega }{dt}=\frac{\partial }{\partial t}\Delta \psi +\left[ \left( -%
\mathbf{\nabla }\psi \times \widehat{\mathbf{e}}_{z}\right) \cdot \mathbf{%
\nabla }\right] \Delta \psi =0.  \label{eq1}
\end{equation}

In $2D$ there is flow of energy in the spectrum from small spatial scales
toward the large spatial scales, \emph{i.e.} inverse cascade. The numerical
simulations confirm this behavior \cite{Montgomery1991}, \cite%
{Montgomery1992}. Adding just a small viscosity and starting from a state of
turbulence, the fluid evolves to a state of highly ordered flow: the
positive and negative vorticities contained in the initial flow are
separated and collected into two large scale vortical flows of opposite
sign. The streamfunction $\psi $ in these states reached asymptotically at
relaxation from turbulence verifies the \emph{sinh}-Poisson equation%
\begin{equation}
\Delta \psi +\lambda \sinh \psi =0  \label{eq2}
\end{equation}%
where $\lambda >0$ is a parameter. This equation is exactly integrable \cite%
{TingChenLee}.

The discretized form of Eq.(\ref{eq1}) has been extensively studied \cite%
{KraichnanMontgomery}, \cite{RobertSommeria}, \cite{RobertSommeria1992}, 
\cite{Chavanis1}. We just review few elements of this theory, for further
reference.

Consider the discretization of the vorticity field $\omega \left( x,y\right) 
$ in a set of $2N$ point-like vortices $\omega _{i}$ each carrying the
elementary quantity $\omega _{0}$ (= const $>0$) of vorticity which can be
positive or negative $\omega _{i}=\pm \omega _{0}$. There are $N$ vortices
with the vorticity $+\omega _{0}\ \ $and $N$ vortices with the vorticity $%
-\omega _{0}$. The current position of a point-like vortex is $\left(
x_{i},y_{i}\right) $ at the moment $t$. The vorticity is expressed as%
\begin{equation}
\omega \left( x,y\right) =\sum\limits_{i=1}^{2N}\omega _{i}a^{2}\delta
\left( x-x_{i}\right) \delta \left( y-y_{i}\right)  \label{omeg}
\end{equation}%
where $a$ is the radius of an effective support of a smooth representation
of the Dirac $\delta $ functions approximating the product of the two $%
\delta $ functions \cite{KraichnanMontgomery}. Then $\omega _{i}a^{2}$
approximates the \emph{circulation} $\gamma _{i}$ which is the integral of
the vorticity over a small area around the point $\left( x_{i},y_{i}\right) $%
: $\gamma _{i}=\int d^{2}x\omega _{i}$ \cite{Chavanis1}. The formal solution
of the equation $\Delta \psi =\omega $, connecting the vorticity and the
streamfunction, can be obtained using the Green function for the Laplace
operator%
\begin{equation}
\Delta _{x,y}G\left( x,y;x^{\prime },y^{\prime }\right) =\delta (x-x^{\prime
})\delta \left( y-y^{\prime }\right)  \label{greendef}
\end{equation}%
where $\left( x^{\prime },y^{\prime }\right) $ is a reference point in the
plane. As shown in Ref.\cite{KraichnanMontgomery} $G\left( \mathbf{r};%
\mathbf{r}^{\prime }\right) $ can be approximated for $a$ small compared to
the space extension of the fluid, $L$, $a\ll L$, as the Green function of
the Laplacian 
\begin{equation}
G\left( \mathbf{r};\mathbf{r}^{\prime }\right) \approx \frac{1}{2\pi }\ln
\left( \frac{\left\vert \mathbf{r}-\mathbf{r}^{\prime }\right\vert }{L}%
\right)  \label{Green}
\end{equation}%
where $L$ is the length of the side of the square domain in plane. The
solution of the equation $\Delta \psi =\omega $ is 
\begin{equation}
\psi \left( \mathbf{r}\right) =\sum\limits_{i=1}^{2N}\gamma _{i}\frac{1}{%
2\pi }\ln \left( \frac{\left\vert \mathbf{r}-\mathbf{r}_{i}\right\vert }{L}%
\right)  \label{psilog}
\end{equation}%
The velocity of the $k$-th point-vortex is $\mathbf{v}_{k}=-\left. \mathbf{%
\nabla }\psi \right\vert _{\mathbf{r=r}_{k}}\times \widehat{\mathbf{e}}_{z}$
and the equations of motion are%
\begin{eqnarray}
\frac{dx_{k}}{dt} &=&v_{x}^{\left( k\right) }=-\sum\limits_{i=1,i\neq
k}^{2N}\gamma _{i}\frac{1}{2\pi }\frac{y_{k}-y_{i}}{\left\vert \mathbf{r}%
_{k}-\mathbf{r}_{i}\right\vert ^{2}}  \label{statiseqs} \\
\frac{dy_{k}}{dt} &=&v_{y}^{\left( k\right) }=\sum\limits_{i=1,i\neq
k}^{2N}\gamma _{i}\frac{1}{2\pi }\frac{x_{k}-x_{i}}{\left\vert \mathbf{r}%
_{k}-\mathbf{r}_{i}\right\vert ^{2}}\   \notag
\end{eqnarray}

The equations are purely kinematic (no inertia), can be derived from a
Hamiltonian and the continuum limit of the discretization is mathematically
equivalent with the fluid dynamics. The standard way of describing the
discrete model is within a statistical approach \cite{Onsager}, \cite%
{KraichnanMontgomery}, \cite{RobertSommeria}, \cite{MajdaWang}. The
elementary vortices are seen as elements of a system of interacting
particles (a gas) that explores an ensemble of microscopic states leading to
the macroscopic manifestation that is the fluid flow. The number of positive
vortices in the state $i$ is $N_{i}^{+}$ and the number of negative vortices
in the state $i$ is $N_{i}^{-}$. The total numbers of positive and
respectively negative vortices are equal: $N^{+}=\sum\limits_{i}N_{i}^{+}=%
\sum\limits_{i}N_{i}^{-}=N^{-}$. This system has a statistical temperature
that is negative when the energy is zero or positive \cite{joycemontgomery}, 
\cite{EdwardsTaylor}. The energy of the discrete system of point-like
vortices is $\mathcal{E}=\frac{1}{2}\sum\limits_{ij}\omega \left( \mathbf{r}%
_{i}\right) G\left( \mathbf{r}_{i},\mathbf{r}_{j}\right) \omega \left( 
\mathbf{r}_{j}\right) $ where $\omega \left( \mathbf{r}_{i}\right) =-\left(
N_{i}^{+}-N_{i}^{-}\right) $ is the vorticity. The probability of a state is
calculated as a combinatorial expression%
\begin{equation}
\mathcal{W}=\left\{ \frac{N^{+}!}{\prod\limits_{i}N_{i}^{+}!}\right\}
\left\{ \frac{N^{-}!}{\prod\limits_{i}N_{i}^{-}!}\right\}   \label{W}
\end{equation}%
The \emph{entropy} is the logarithm of this expression and by extremization
one finds 
\begin{equation}
\ln N_{i}^{\pm }+\alpha ^{\pm }\pm \beta \sum\limits_{j}G\left( \mathbf{r}%
_{i},\mathbf{r}_{j}\right) \left( N_{j}^{+}-N_{j}^{-}\right) =0  \label{lnn}
\end{equation}%
for $i=1,N$, where $\alpha ^{\pm }$ and $\beta $ are Lagrange multipliers
introduced to ensure $\sum N_{i}^{+}=\sum N_{i}^{-}=N=$ const and
conservation of the Energy $\mathcal{E}$. The solutions are written in terms
of a continuous function $\psi \left( x,y\right) $%
\begin{equation}
N_{i}^{\pm }=\exp \left[ -\alpha ^{\pm }\mp \beta \psi \left( x,y\right) %
\right]   \label{npsi}
\end{equation}%
implying $N_{i}^{+}N_{i}^{-}=$ const. From Eq.(\ref{lnn}) and (\ref{npsi})
the \emph{sinh}-Poisson equation (\ref{eq2}) is derived.

\section{Field theoretical formulation of the continuum limit of the
point-like vortex model}

The model of point-like vortices captures the physics of the $2D$ Euler
fluid in a new formulation: matter (density of point-like vortices), field
(the Coulombian potential in Eq.(\ref{psilog}) ) and interaction. This
suggests to formulate the continuum limit of the discrete point-like
vortices as a field theory. The density of point-like vortices is
represented by the matter field $\phi \left( x,y,t\right) $ and the
potential of interaction by the gauge field $A_{\mu }\left( x,y,t\right) $, $%
\mu =0,1,2$. \ The \textquotedblleft matter\textquotedblright\ consists of
the positive and negative vortices.The dynamics is $2D$ but we exploit the
invariance to motion along the third axis to reveal the chiral nature of the
elementary vortices. The positive vortices: (1) rotate anti-clockwise in
plane: $\omega \widehat{\mathbf{e}}_{z}\sim \mathbf{\sigma }$\ spin is up;
(2) move along the positive $z$ axis: $\mathbf{p=}\widehat{\mathbf{e}}%
_{z}p_{0}$; (3) have positive chirality: $\chi =\mathbf{\sigma \cdot p/}%
\left\vert \mathbf{p}\right\vert $. The positive vortices can be represented
as a point that runs along a positive helix, upward. In projection from the
above the plane toward the plane we see a circle on which the point moves
anti-clockwise.

The \ negative vortices: (1) rotate clockwise in plane: $\left( -\omega
\right) \widehat{\mathbf{e}}_{z}\sim -\mathbf{\sigma }$ spin is down; (2)
move along the negative $z$ axis: $-\mathbf{p=}\widehat{\mathbf{e}}%
_{z}\left( -p_{0}\right) $, along $-z$; (3) have positive chirality: $\chi =%
\mathbf{\sigma \cdot p/}\left\vert \mathbf{p}\right\vert $. The negative
vortices can be represented as a point that runs along a positive helix, the
same as above, but runs downward. In projection from the above the plane
toward the plane we see a circle on which the point moves clockwise.

The positive vortices and the negative vortices have the same \emph{%
chirality }and in a point where there is superposition of a positive and a
negative elementary vortices the \emph{chirality} is added. In particular,
the vacuum consists of paired positive and negative vortices, with no motion
of the fluid, which in physical variables means $\psi \equiv 0$, $\mathbf{%
v\equiv 0}$, $\omega \equiv 0$. In FT the vacuum consists of superposition
of positive and negative vortices, which means: (1) zero spin, or zero \emph{%
vorticity}; (2) zero momentum $\mathbf{p}=\mathbf{0}$; (3) $2\times $chiral
charge. The Euler fluid at equilibrium ($\psi =0$, $\mathbf{v=0}$, $\omega
=0 $) is in a vacuum with \emph{broken chiral invariance}.

The $sl(2,\mathbf{C})$ Non-Abelian structure is necessary due to the \emph{%
vortical} nature of the elementary object: the vorticity matter must be
represented by a \emph{mixed spinor}. The Lagrangian \cite{dunnebook}, \cite%
{jackiwpi}%
\begin{eqnarray}
\mathcal{L} &=&-\kappa \varepsilon ^{\mu \nu \rho }\mathrm{tr}\left( \left(
\partial _{\mu }A_{\nu }\right) A_{\rho }+\frac{2}{3}A_{\mu }A_{\nu }A_{\rho
}\right)  \label{Lagrange} \\
&&+i\mathrm{tr}\left( \phi ^{\dagger }\left( D_{0}\phi \right) \right) -%
\frac{1}{2m}\mathrm{tr}\left( \left( D_{k}\phi \right) ^{\dagger }\left(
D^{k}\phi \right) \right) \   \notag \\
&&+\frac{1}{4m\kappa }\mathrm{tr}\left( \left[ \phi ^{\dagger },\phi \right]
^{2}\right)  \notag
\end{eqnarray}%
where $D_{\mu }=\partial _{\mu }+\left[ A_{\mu },\right] $ and $\kappa $, $m$
are positive constants. The Euler - Lagrange equations for the action
functional $\mathcal{S}=\int dxdydt\mathcal{L}$ are the equations of motion 
\begin{eqnarray}
iD_{0}\phi &=&-\frac{1}{2m}D_{k}D^{k}\phi -g\left[ \left[ \phi ,\phi
^{\dagger }\right] ,\phi \right]  \label{eqmotion} \\
\kappa \varepsilon ^{\mu \nu \rho }F_{\nu \rho } &=&iJ^{\mu }\   \notag
\end{eqnarray}%
where the current $J^{\mu }=\left( J^{0},J^{k}\right) $ 
\begin{eqnarray}
J^{0} &=&\left[ \phi ,\phi ^{\dagger }\right]  \label{current} \\
J^{k} &=&-\frac{i}{2m}\left( \left[ \phi ^{\dagger },\left( D^{k}\phi
\right) \right] -\left[ \left( D^{k}\phi \right) ^{\dagger },\phi \right]
\right) \   \notag
\end{eqnarray}%
is covariantly conserved $D_{\mu }J^{\mu }=0$. The energy density is 
\begin{equation}
E=\frac{1}{2m}\mathrm{tr}\left( \left( D_{k}\phi \right) ^{\dagger }\left(
D^{k}\phi \right) \right) -\frac{g}{2}\mathrm{tr}\left( \left[ \phi
^{\dagger },\phi \right] ^{2}\right)  \label{energia}
\end{equation}

The Gauss constraint is the zero component of the second equation of motion 
\begin{equation}
2\kappa F_{12}=iJ^{0}=i\left[ \phi ,\phi ^{\dagger }\right]  \label{ga1}
\end{equation}

In the following we will use the combinations: $A_{\pm }\equiv A_{x}\pm
iA_{y}$, $\partial /\partial z=\frac{1}{2}\left( \partial /\partial
x-i\partial /\partial y\right) $, $\partial /\partial z^{\ast }=\frac{1}{2}%
\left( \partial /\partial x+i\partial /\partial y\right) $, and similar.
Writing 
\begin{eqnarray}
\mathrm{tr}\left( \left( D_{k}\phi \right) ^{\dagger }\left( D^{k}\phi
\right) \right)  &=&\mathrm{tr}\left( \left( D_{-}\phi \right) ^{\dagger
}\left( D_{-}\phi \right) \right) -i\mathrm{tr}\left( \phi ^{\dagger }\left[
F_{12},\phi \right] \right)   \label{dkdk} \\
&&-m\varepsilon ^{ij}\partial _{i}\left[ \phi ^{\dagger }\left( D_{j}\phi
\right) -\left( D_{j}\phi \right) ^{\dagger }\phi \right] \   \notag
\end{eqnarray}%
we replace in the expression Eq.(\ref{energia}) of the energy density and
note that for smooth fields we can ignore the last term, which is evaluated
at the boundary 
\begin{equation}
E=\frac{1}{2m}\mathrm{tr}\left( \left( D_{-}\phi \right) ^{\dagger }\left(
D_{-}\phi \right) \right)   \label{enersd}
\end{equation}%
The states are \emph{static} $\partial _{0}\phi =0$ and minimize the energy (%
$E=0$). Adding the Gauss constraint (after replacing $F_{12}=\left(
i/2\right) F_{+-}$) we have a set of two equations for stationary states
corresponding to the absolute minimum of the energy%
\begin{eqnarray}
D_{-}\phi  &=&0  \label{eqmotsd} \\
F_{+-} &=&\frac{1}{\kappa }\left[ \phi ,\phi ^{\dagger }\right] \   \notag
\end{eqnarray}

From these equations the \emph{sinh}\textbf{-}Poisson equation is derived 
\cite{DunneJackiwTrugenberg}, \cite{florinmadi2003}. The states correspond
to zero curvature in a formulation that involves the reduction to $2D$ from
a four dimensional Self - Dual Yang Mills system, as shown in \cite%
{DunneJackiwTrugenberg}. Therefore we will denote this state as Self - Dual
(SD). The functions $\phi $ and $A_{\mu }$ are mixed spinors, elements of
the algebra $sl\left( 2,\mathbf{C}\right) $.

In order to solve this system and connect with variables of the real fluid,
one can adopt the following algebraic ansatz \cite{dunnebook},%
\begin{equation}
\phi =\phi _{1}E_{+}+\phi _{2}E_{-}\ ,\ \phi ^{\dagger }=\phi _{1}^{\ast
}E_{-}+\phi _{2}^{\ast }E_{+}  \label{ansatz1}
\end{equation}%
and%
\begin{equation}
A_{-}=aH\ ,\ A_{+}=-a^{\ast }H  \label{ansatz2}
\end{equation}%
which is based on the three generators $\left( E_{+},H,E_{-}\right) $ of the
Chevalley basis of $sl\left( 2,\mathbf{C}\right) $. Explicitly: $%
E_{+}=\left( 
\begin{array}{cc}
0 & 1 \\ 
0 & 0%
\end{array}%
\right) $, $E_{-}=\left( 
\begin{array}{cc}
0 & 0 \\ 
1 & 0%
\end{array}%
\right) $ and $H=\left( 
\begin{array}{cc}
1 & 0 \\ 
0 & -1%
\end{array}%
\right) $. From the first equation of motion $D_{-}\phi =0$ we obtain 
\begin{equation}
\frac{\partial \phi _{1}}{\partial z}+a\phi _{1}=0  \label{64a}
\end{equation}%
\begin{equation}
\frac{\partial \phi _{2}}{\partial z}-a\phi _{2}=0  \label{64b}
\end{equation}%
The Gauss equation becomes

\begin{equation}
\frac{\partial a}{\partial x_{+}}+\frac{\partial a^{\ast }}{\partial x_{-}}=%
\frac{1}{k}\left( \rho _{1}-\rho _{2}\right)  \label{56}
\end{equation}%
where $\rho _{1,2}\equiv \left\vert \phi _{1,2}\right\vert ^{2}$. Using Eqs.(%
\ref{64a}) and its complex conjugate the left hand side of Eq.(\ref{56})
becomes%
\begin{equation*}
\frac{\partial a}{\partial x_{+}}+\frac{\partial a^{\ast }}{\partial x_{-}}%
=-2\frac{\partial ^{2}}{\partial z\partial z^{\ast }}\ln \left( \left\vert
\phi _{1}\right\vert ^{2}\right) =-\frac{1}{2}\Delta \ln \left( \left\vert
\phi _{1}\right\vert ^{2}\right)
\end{equation*}%
or 
\begin{equation}
-\frac{1}{2}\Delta \ln \rho _{1}=\frac{1}{\kappa }\left( \rho _{1}-\rho
_{2}\right)  \label{82}
\end{equation}%
Similarly, using Eq.(\ref{64b}) in Eq.(\ref{56}) we obtain 
\begin{equation*}
\frac{\partial a}{\partial x_{+}}+\frac{\partial a^{\ast }}{\partial x_{-}}=2%
\frac{\partial ^{2}}{\partial z\partial z^{\ast }}\ln \left( \left\vert \phi
_{2}\right\vert ^{2}\right) =\frac{1}{2}\Delta \ln \left( \left\vert \phi
_{2}\right\vert ^{2}\right)
\end{equation*}%
or 
\begin{equation}
\frac{1}{2}\Delta \ln \rho _{2}=\frac{1}{\kappa }\left( \rho _{1}-\rho
_{2}\right)  \label{86}
\end{equation}%
The right hand side in Eqs.(\ref{82}) and (\ref{86}) is the same and if we
substract the equations we obtain 
\begin{equation}
\Delta \ln \rho _{1}+\Delta \ln \rho _{2}=0  \label{87}
\end{equation}%
It results that $\rho _{1}=\rho _{2}^{-1}\equiv \rho $. Now we introduce a
scalar function $\psi $, defined by $\rho =\exp \left( \psi \right) $ and
the Eqs.(\ref{82}) and (\ref{86}) take the unique form%
\begin{equation}
\Delta \ln \rho =-\frac{2}{\kappa }\left( \rho -\frac{1}{\rho }\right)
\label{eqbun}
\end{equation}%
which is the \emph{sinh}-Poisson equation (also known as the elliptic \emph{%
sinh}-Gordon equation)%
\begin{equation}
\Delta \psi +\frac{4}{\kappa }\sinh \psi =0  \label{sinhp}
\end{equation}

The states identified by the FT are the absolute extrema of the action
functional and are characterized by: (1) stationarity; (2) double
periodicity, \emph{i.e.} the function $\psi \left( x,y\right) $ must only be
determined on a \textquotedblleft fundamental\textquotedblright\ square in
plane; (3) the total vorticity is zero; (4) the states verify Eq.(\ref{sinhp}%
). A more detailed discussion can be found in \cite{florinmadi2003}, \cite%
{florinmadiXXX2013}, \cite{flmadichaos}. The field theoretical model is a
reformulation of the system of point-like vortices. The parallel between the
two formulations has led to the conclusion that the extremum of entropy (for
negative statistical temperature) corresponds to the states of self-duality.

\section{The surface in the Euclidean $\mathbf{E}^{3}$ space}

It is interesting that the same equation governing the asymptotic states of
the $2D$ Euler fluid is also the equation that identifies the surfaces in $%
\mathbf{E}^{3}$ that have Constant Mean Curvature (CMC). To make more clear
the connection we review very briefly the theory of Constant Mean Curvature
(CMC) surfaces, following \cite{bobenkouspekhi}.

A surface is a mapping from a domain in the plane $\left( x,y\right) $ to
the points $\mathbf{F}=\left( F_{1},F_{2},F_{3}\right) \in \mathbf{E}^{3}$.
Equivalently, one can use complex variables $\left( x,y\right) \rightarrow
\left( z,z^{\ast }\right) $ giving the conformal parametrization of the
surface and the metric $ds^{2}=\exp \left( u\right) \left(
dx^{2}+dy^{2}\right) =\exp \left( u\right) dzdz^{\ast }$ where $u\left(
z,z^{\ast }\right) $ is a real function. The following normalizations are
applied to the vectors in the tangent plane in the current point of the
surface%
\begin{equation}
\frac{\partial \mathbf{F}}{\partial z}\cdot \frac{\partial \mathbf{F}}{%
\partial z^{\ast }}=\frac{1}{2}\exp \left( u\right) \ ,\ \frac{\partial 
\mathbf{F}}{\partial z}\cdot \frac{\partial \mathbf{F}}{\partial z}=0\ \ ,\
\ \frac{\partial \mathbf{F}}{\partial z^{\ast }}\cdot \frac{\partial \mathbf{%
F}}{\partial z^{\ast }}=0  \label{s1}
\end{equation}%
The vector normal to the tangent plane is $\mathbf{N\sim }\frac{\partial 
\mathbf{F}}{\partial z}\times \frac{\partial \mathbf{F}}{\partial z^{\ast }}$
and is normalized $\mathbf{N\cdot N}=1$. The three vectors define the \emph{%
moving frame}%
\begin{equation}
\sigma \equiv \left( \frac{\partial \mathbf{F}}{\partial z},\frac{\partial 
\mathbf{F}}{\partial z^{\ast }},\mathbf{N}\right) ^{T}  \label{s2}
\end{equation}%
The change of a vector of the moving frame at an infinitesimal displacement
on the surface is expressed by Christoffel coefficients. More compactly, the
change of the frame (\emph{Gauss Weingarten }equations) is expressed in
terms of two $3\times 3$ matrices $\left( \mathcal{U},\mathcal{V}\right) $ 
\begin{equation}
\frac{\partial \sigma }{\partial z}=\mathcal{U}\sigma \ ,\ \frac{\partial
\sigma }{\partial z^{\ast }}=\mathcal{V}\sigma  \label{s3}
\end{equation}%
The consistency condition is the equality of mixed derivatives (Gauss
Codazzi equations) $\frac{\partial \mathcal{U}}{\partial z^{\ast }}-\frac{%
\partial \mathcal{V}}{\partial z}+\left[ \mathcal{U},\mathcal{V}\right] =0$.
\ The first form $I=d\mathbf{F\cdot }d\mathbf{F}$ and the second form $II=d%
\mathbf{F\cdot N}$ of the surface are expressed by the formulas%
\begin{equation}
I=\exp \left( u\right) \left( 
\begin{array}{cc}
1 & 0 \\ 
0 & 1%
\end{array}%
\right)  \label{s4}
\end{equation}%
\begin{equation}
II=\left( 
\begin{array}{cc}
Q+Q^{\ast }+H\exp \left( u\right) & i\left( Q-Q^{\ast }\right) \\ 
i\left( Q-Q^{\ast }\right) & -\left( Q+Q^{\ast }\right) +H\exp \left(
u\right)%
\end{array}%
\right)  \label{s5}
\end{equation}%
in terms of the projections of the second order derivatives of $\mathbf{F}$
on the normal versor%
\begin{equation}
\frac{\partial ^{2}\mathbf{F}}{\partial z\partial z}\cdot \mathbf{N=}Q\ ,\ 
\frac{\partial ^{2}\mathbf{F}}{\partial z\partial z^{\ast }}\cdot \mathbf{N}=%
\frac{1}{2}H\exp \left( u\right)  \label{s6}
\end{equation}

The \emph{principal curvatures} $\kappa _{1}$ and $\kappa _{2}$ are
eigenvalues of the matrix $\left( II\right) \cdot I^{-1}$ and the mean - ,
respectively the Gauss curvature are $H=\frac{1}{2}\left( \kappa _{1}+\kappa
_{2}\right) $ and $K=\kappa _{1}\kappa _{2}=H^{2}-4QQ^{\ast }\exp \left(
-2u\right) $. The \emph{Gauss-Codazzi} equations become%
\begin{eqnarray}
\frac{\partial ^{2}u}{\partial z\partial z^{\ast }}+\frac{1}{2}H^{2}\exp
\left( u\right) -2QQ^{\ast }\exp \left( -u\right)  &=&0  \label{s7} \\
\frac{\partial Q}{\partial z^{\ast }} &=&\frac{1}{2}\frac{\partial H}{%
\partial z}\exp \left( u\right)   \notag
\end{eqnarray}%
If the mean curvature is constant: $H=1/2$ the function $Q$ is holomorphic
and can be taken constant $\neq 0$ (assuming that there is no umbilic
point). Taking $Q=1/4$ the first equation becomes the \emph{sinh}-Poisson
equation $\Delta u+\sinh \left( u\right) =0$. To make it coincide with (\ref%
{sinhp}) one chooses the constants: $H=1/\sqrt{\kappa }$ , $Q=1/\left( 2%
\sqrt{\kappa }\right) $ where $\kappa $ is the coefficient of the
Chern-Simons term in (\ref{Lagrange}). For any flow in the asymptotic regime
of the Euler fluid, there is a corresponding CMC surface. From point-like
vortices and from FT models we know that the solutions of sinh-Poisson for
fluids must be doubly periodic, which means that the CMC surfaces must be
tori. We identify $u\rightarrow \psi $ (the streamfunction) and $\rho =\exp
\psi $ is the factor in the isothermal metric Eq.(\ref{s4}) of the surface 
\cite{bobenkouspekhi}. The Gaussian curvature is $K=-\omega /\left( 2\rho
\right) =\kappa ^{-1}\left( 1-1/\rho ^{2}\right) $ where $\omega $ is the
vorticity. The integral on the surface of the Gaussian curvature is $\iint
KdA=2\pi \chi =0$, where $\chi $ is the Euler characteristic of the torus.
This was expected because $\iint KdA=\iint K\exp \left( \psi \right) dxdy=-%
\frac{1}{2}\iint \omega dxdy=0$, since we know that the absolute extremum
for both models SPV and FT finds zero total vorticity in the domain.

This purely geometric derivation of the \emph{sinh}-Poisson equation seems
far from the other two discussed above: the extremum of entropy for the
statistical ensemble of point-like vortices and the extremum of the action
functional in the field theoretical model. However the CMC surfaces can also
be derived from an extremum condition. The mean curvature $H=\left(
1/R_{1}+1/R_{2}\right) /2$ of a surface ($R_{1,2}$ are the radii along
principal curvature lines) intervenes in the balance between the capillary
force and the difference $\Delta p$ of pressure on the two sides, the
Laplace law: $\Delta p=2\sigma H$ where $\sigma =$const. is the coefficient
of surface tension. The CMC surface is \textquotedblleft in mechanical
equilibrium\textquotedblright\ since $H=$ constant means equal pressure $%
\Delta p$ in all points. For surfaces the principle of extremum is connected
with the energy of the capillary forces, $E=\sigma A$, where $A$ is the area
of the surface. It is clear that the \textquotedblleft
equilibrium\textquotedblright\ mentioned above corresponds to the minimum
energy, \emph{i.e.} of the area $A$. Then the \emph{sinh}-Poisson equation
appears again as being derived from a principle of extremum. Explicit
analytical application of the minimization of surface at fixed volume can be
done for the unduloids of Delaunay \cite{oprea1}, well-known CMC surfaces.

Regarding the asymptotic states we note that to every flow ($u\equiv $
theta-function solution of sinh-Poisson equation) one can now associate a
CMC surface with metric (\ref{s4}). The reverse direction, \emph{i.e.} to
derive the physical possibility of a stationary flow from particularities of
the CMC surfaces (like embedding versus immersion, multiple ends, etc.) has
been much less investigated.

Random perturbations on the surface, violating the CMC condition,
corresponds to the regime that precedes the asymptotic ordered states of the
fluid. The tendency of capillary forces to smooth out the perturbations and
to install the CMC state, corresponds to the fluid being driven by inverse
cascade to the coherent dipolar flow, and with the extremization of the FT
action leading to the self-dual states. The local perturbations of the
surface, by which it departs from CMC $H\neq $ const, are corrected by
curvature flow, where the velocity of a point of the surface along the
normal is proportional to the local mean curvature, a well known dynamics of
interfaces in physical systems. In the model of point-like vortices the
entropy decreases and the temperature is negative when the system evolves to
organization \cite{joycemontgomery}. Similarly, the entropy, defined for a
line as $S=\alpha \int ds\ H\left( t\right) \ln H\left( t\right) $,
decreases when the curvature $H\left( t\right) $ evolves according to the
curvature flow \cite{curvflowzhu}.

For $H=1/\sqrt{\kappa }$ and $K=\kappa ^{-1}\left( 1-1/\rho ^{2}\right) $ it
results $\left( \sqrt{\kappa }\rho \right) ^{-1}=\kappa _{2}-\kappa _{1}$.
If an umbilic point $\left( \kappa _{1}=\kappa _{2}\right) $ existed on the
surface, this would correspond to a singular vorticity $\rho \rightarrow
\infty $. Since the CMC surface associated to the fluid flow is a torus an
umbilic point does not exist which means that a singular vorticity cannot
exist in the asymptotic state of the flow. But in the regime that precedes
it we may expect that a localized, high concentration of vorticity should
correspond on the perturbed quasi-CMC surface to a position where there is
approximate equality of the principal curvatures.

\section{The chiral model and the uniton}

The equations of self-duality of the FT model, (\ref{eqmotsd}) can be
formulated as equations of the chiral model. The suggestion comes from the
observation that the new potential defined by $\mathcal{A}_{-}=A_{-}+\sqrt{%
\frac{1}{\kappa }}\phi ^{\dagger }$ and $\mathcal{A}_{+}=A_{+}-\sqrt{\frac{1%
}{\kappa }}\phi $ has zero curvature \cite{dunnebook}%
\begin{equation}
\mathcal{F}_{+-}=\partial _{+}\mathcal{A}_{-}-\partial _{-}\mathcal{A}_{+}+%
\left[ \mathcal{A}_{+},\mathcal{A}_{-}\right] =0  \label{u3}
\end{equation}%
if $A_{\pm }$ and $\phi $ verify the SD equations (\ref{eqmotsd})
(independently of the algebraic ansatz; if the algebraic ansatz is adopted \
the equation $\mathcal{F}_{+-}=0$ leads to the \emph{sinh}-Poisson
equation). Then the new potential is locally a gauge field $\mathcal{A}%
_{-}=g^{-1}\partial _{-}g$. Using $g$ to transform the matter field $\phi $ 
\begin{equation}
\chi =\sqrt{\frac{1}{\kappa }}g\phi g^{-1}  \label{u4}
\end{equation}%
it can be checked \cite{dunnebook} that the two SD (\ref{eqmotsd}) equations
are verified if $\chi $ is the solution of%
\begin{equation}
\partial _{-}\chi =\left[ \chi ^{\dagger },\chi \right]   \label{u5}
\end{equation}%
Then, expressing $\chi $ as%
\begin{equation}
\chi \equiv \frac{1}{2}h^{-1}\partial _{+}h  \label{u6}
\end{equation}%
one obtains the equation of the chiral model \cite{manas}, \cite{wardsigma}%
\begin{equation}
\partial _{+}\left( h^{-1}\partial _{-}h\right) +\partial _{-}\left(
h^{-1}\partial _{+}h\right) =0  \label{u7}
\end{equation}%
where $h$ is a map from a domain in the Euclidean plane to the group $%
SL\left( 2,\mathbf{C}\right) $. Until this point no special assumption is
made on the algebraic content of $h$ (hence of $\phi $, through $\chi $).
Finding a solution of this equation provides us with a more general solution
than that offered by the \emph{sinh}-Poisson equation, since the latter is
derived under the simplified algebraic ansatz, Eqs.(\ref{ansatz1} - \ref%
{ansatz2}). The solutions of the chiral model Eq.(\ref{u7}) are unitons, 
\emph{i.e.} harmonic maps $\mathbf{R}^{2}\rightarrow SL\left( 2,\mathbf{C}%
\right) $. For this group the construction of uniton is based on a single
rational function $f\left( z\right) $ \cite{uhlenbeckunitons}, \cite%
{zakrzewski1}, \cite{dunnebook} and has the expression \cite{wardunitons} 
\begin{equation}
h=\frac{i}{1+\left\vert f\right\vert ^{2}}\left( 
\begin{array}{cc}
1-\left\vert f\right\vert ^{2} & f \\ 
f^{\ast } & -1+\left\vert f\right\vert ^{2}%
\end{array}%
\right)   \label{u8}
\end{equation}%
\begin{equation}
\chi =\frac{1}{2}h^{-1}\partial _{+}h=\frac{f}{\left( 1+\left\vert
f\right\vert ^{2}\right) ^{2}}2\frac{\partial f^{\ast }}{\partial z^{\ast }}%
\left( 
\begin{array}{cc}
1 & f \\ 
-\frac{1}{f} & -1%
\end{array}%
\right)   \label{u9}
\end{equation}%
All Chevalley generators are present. Calculating the matrix $\left[ \chi
,\chi ^{\dagger }\right] $ it is shown \cite{dunnebook} that this can be
diagonalized by a unitary matrix $g$ and using Eq.(\ref{u4}) one obtains the 
\emph{covariant charge density}, Eq.(\ref{current})%
\begin{eqnarray}
J^{0} &=&\left[ \phi ,\phi ^{\dagger }\right] =\kappa g^{-1}\left[ \chi
,\chi ^{\dagger }\right] g  \label{u12} \\
&=&-\kappa \partial _{+}\partial _{-}\ln \left( 1+\left\vert f\right\vert
^{2}\right) \left( 
\begin{array}{cc}
1 & 0 \\ 
0 & -1%
\end{array}%
\right)   \notag
\end{eqnarray}%
The fact that the matter density $\left[ \phi ,\phi ^{\dagger }\right] $ has
only the Cartan component $H$ may be seen as justifying the ansatz $A_{\pm
}\sim H$. As we know, applying the Eqs.(\ref{ansatz1}), (\ref{ansatz2}) to
the SD equations (\ref{eqmotsd}) and (\ref{eqbun}) it results 
\begin{equation}
\left[ \phi ,\phi ^{\dagger }\right] =\left( \rho -\frac{1}{\rho }\right)
H=\left( -\frac{\kappa }{2}\Delta \ln \rho \right) H  \label{u13}
\end{equation}%
and there is the correspondence $1+\left\vert f\right\vert ^{2}=\exp \left(
\psi /2\right) =\sqrt{\rho }$. Here we recall the matrix $\Phi $ that
transforms the euclidean frame $\left( \widehat{\mathbf{e}}_{1}+i\widehat{%
\mathbf{e}}_{2},\widehat{\mathbf{e}}_{1}-i\widehat{\mathbf{e}}_{2},\widehat{%
\mathbf{e}}_{3}\right) $ into the moving frame $\left( \frac{\partial 
\mathbf{F}}{\partial z},\frac{\partial \mathbf{F}}{\partial z^{\ast }},%
\mathbf{N}\right) $ attached to the current point of the surface, by the
formulas \cite{bobenkomathann}%
\begin{eqnarray}
\frac{\partial F}{\partial z} &=&-i\exp \left( \frac{u}{2}\right) \Phi
^{-1}E_{-}\Phi   \label{u17} \\
\frac{\partial F}{\partial z^{\ast }} &=&-i\exp \left( \frac{u}{2}\right)
\Phi ^{-1}E_{+}\Phi   \notag
\end{eqnarray}%
These expressions are used in the Gauss-Weingarten equations (\ref{s3}). The
compatibility condition, the definitions Eq.(\ref{s6}) and the normalization 
$\det \Phi =\exp \left( u/2\right) $ determine explicitly $\Phi $ \cite%
{bobenkoexploint}. Comparing $1+\left\vert f\right\vert ^{2}=\exp \left(
\psi /2\right) $ with $\det \Phi =\exp \left( \frac{\psi }{2}\right) $ we
are suggested that $1+\left\vert f\right\vert ^{2}$ has for unitons the
similar meaning as $\det \left( \Phi \right) $ for CMC surfaces, with $\Phi $
the \emph{quaternion} which provides the mapping between (a) the euclidean
fixed orthogonal frame, to (b) the moving frame attached to the current
point $\mathbf{F}\equiv F$ of the surface. Indeed, the equation (\ref{u7})
of the chiral model is equivalent to a system (Weierstrass-Enneper) of
equations defining CMC surfaces \cite{konopelchenkotaimanov} . The uniton (%
\ref{u8}) of the chiral model is also obtained in terms of solutions of
these equations for constant mean curvature surfaces \cite{wei3}.

The energy of the chiral model \cite{dunnebook} \cite{wardunitons} is given
by the abelian charge of the FT 
\begin{equation*}
E\left[ h\right] =\frac{1}{2}\int d^{2}x\ \mathrm{Tr}\left[ \left(
h^{-1}\partial _{-}h\right) \left( h^{-1}\partial _{+}h\right) \right] =%
\frac{2}{\kappa }Q^{A}
\end{equation*}%
The density of $Q^{A}$ is $\mathrm{Tr}\left( \phi ^{\dagger }\phi \right)
=\rho +1/\rho $. The fact that the energy of the chiral model and the
abelian charge $Q^{A}$ are integer multiples of some physical constant is
connected in our case with the property of double periodicity of the
absolute extremum of the FT action, as for entropy in SPV. From those models
we know that for Eq.(\ref{eq2}) one can only retain solutions expressed in
terms of Riemann theta functions.

In summary, the chiral model with equation (\ref{u7}) is an equivalent form
of the FT model at self-duality, with equations Eqs.(\ref{eqmotsd}) and is
also equivalent to the constant mean curvature surface model, with the
Weierstrass-Enneper equations for the matrix $\Phi $.

\section{A set of time-dependent unitons}

The physical system in the regime preceding the static final state consists
of few large-amplitude vortices\ moving slowly in a weak turbulent field.
The unitons Eq.(\ref{u8}) constitute a possible representation of the
quasi-coherent structures (lumps of energy) but they are static since they
solve the SD equations (\ref{eqmotsd}) or equivalently Eq.(\ref{u7}). We
should now leave the asymptotic state and use the equations of motion with
time dependence (\ref{eqmotion}) at least in its close proximity. One should
use the procedure of Manton \cite{manton} as applied, for example, to the
Abelian Higgs (AH) model of superconductivity. The exact solutions of the AH
self-duality equations are static localized vortices (topological defects)
parametrized by the positions of their centers. All possible sets of
parameters constitutes a manifold, the moduli space of SD solutions.
Assuming now that the parameters can be slow functions of time, the
solutions are inserted in the Lagrangian. The resulting structure of the
Lagrangian induces a metric on the manifold and the part which is quadratic
in the time derivatives drives the motion of the vortices, as a mechanical
kinetic energy. It is a geodesic flow on the manifold.

For this procedure to be applied to the SD or chiral solutions it has been
first necessary to modify the uniton equation by extending it with time
dependence \cite{wardunitons}. The derivation operators in Eq.(\ref{u6}) are
replaced $\partial _{\pm }\rightarrow \partial _{\pm }\pm i\partial _{t}$
and the equation which replaces (\ref{u7}) is 
\begin{equation}
\left( \eta ^{\mu \nu }+\varepsilon ^{\mu \nu \rho }V_{\rho }\right)
\partial _{\nu }\left( J^{-1}\partial _{\mu }J\right) =0  \label{t1}
\end{equation}%
where $\eta ^{\mu \nu }=diag\left( -1,1,1\right) $, $\varepsilon ^{\mu \nu
\rho }$ is the totally antisymmetric tensor and $V_{\rho }$ is a fixed
versor with the $\left( t,x,y\right) $ components $\left( 0,1,0\right) $ 
\cite{wardchiral88}. The solution of this new equation has the same form as
the uniton Eq.(\ref{u8}) but now the rational function $f\left( z\right) $
depends on time. Taking%
\begin{equation}
f\left( z\right) =\prod\limits_{k=1}^{N}\left( z-p_{k}\right)
\prod\limits_{m=1}^{N}\left( z-q_{m}\right) ^{-1}  \label{t2}
\end{equation}%
the positions of zero's and poles of $f$ are assumed to be functions of time
with trajectory $\gamma \left( t\right) \equiv \left( p_{k}\left( t\right)
,q_{m}\left( t\right) \right) _{k,m=1,N}$ . Inserting $f$ in Eq.(\ref{u8})
the time variation in the new equation of the uniton $J$, Eq.(\ref{t1}), is
calculated $\partial _{t}J=\frac{\delta J\left[ \gamma \left( t\right) %
\right] }{\delta \gamma _{i}}\frac{d\gamma _{i}\left( t\right) }{dt}$, $%
i=1,2N$. The kinetic terms in the energy-momentum tensor $T=-\frac{1}{2}\int
d^{2}x\mathrm{Tr}\left( J^{-1}\partial _{t}J\right) ^{2}$ take the general
form $h_{jk}\frac{d\gamma _{j}}{dt}\frac{d\gamma _{k}}{dt}$ and the motion
is governed by the constants $h_{jk}$. Explicit calculation of the metric $%
h_{jk}$ of the manifold of uniton solutions are provided in \cite{wardslowly}%
, \cite{dunajskimanton}.

This approach is probably the closest to our objective of studying the
dynamics of structures near the static self-dual state. It will be necessary
to introduce a random factor in the parameters of the unitons, \emph{i.e.}
in the positions of zeros and poles of the rational function $f$. Another
possibility is to follow Dunajski and Manton \cite{dunajskimanton} and
consider that an external magnetic field (which we can take random) alters
the geodesic motion in the manifold which represents the moduli space of
uniton solutions.

\bigskip

\section{Mapping between formulations}

Returning to the connection between the chiral model equation (\ref{u7}) and
the constant mean curvature equations for $\Phi $ one should investigate how
the dynamics of unitons can be mapped to a similar time variation of the
matrix $\Phi $ which defines the moving frame on the surface. This mapping
can be formulated as follows.

Consider a solution $\psi \left( x,y\right) $ of the exactly integrable
equation (\ref{eq2}). It is the streamfunction of a stationary flow of the
Euler fluid. Using the conformal metric (\ref{s4}) and the second form (\ref%
{s5}) with $Q$, $H$ chosen to lead to the same form of sinh-Poisson
equation, one can construct the constant mean curvature surface that
corresponds to that flow. This also gives the expression of the matrix $\Phi 
$, which connects the fixed Euclidean frame to the moving frame on the
surface. All this is strictly limited to the asymptotic state. Now assume
that the CMC surface is perturbed. Random protuberances of low amplitude are
admitted and the exact CMC property is lost. But the CMC property is the
result of a variational extremization of a functional: minimum area for
fixed volume. An evolution similar to the effect of capillarity on an
elastic membrane (foam) will try to return the surface to the CMC state, via
curvature flow. For a single protuberance the curvature flow will dissipate
the area such as to draw the local curvatures $\kappa _{1,2}$ to the values
of the CMC. For a random ensemble of perturbations, the curvature flow may
induce displacements and coalescences as part of the area dissipation. In
any moment of such evolution one can calculate the matrix $\Phi $ since the
frame in any point of the perturbed surface is known. Using $\Phi $ one can
infer the function $f$ \emph{i.e.} the expression of the uniton (\ref{u8}).
When it is represented in $\mathbf{R}^{2}$, the density of energy is a set
of lumps \cite{zakrzewski1}. The evolution will also include changes of
topology: a function $f$ as in (\ref{t2}) induces through (\ref{u8}) a
mapping $\mathbf{R}^{2}\rightarrow S^{2}$ with topological charge \cite%
{wardsigma}. For the perturbation of the surfaces under the curvature flow
the suppression of a perturbation is smooth but a smooth change in $\Phi $
can lead to changes of $f$ across topological classes.

This program is analytically complicated but it offers a possible comparison
with experiment. The rate of return of the perturbed surface to the CMC
state through area dissipation governed by curvature flow will translate
into a rate of change in the portrait of lumps in the density of energy of
the uniton field. Their number should decrease, since the evolution should
get closer to the dipolar structure, solution of sinh-Poisson equation. But
this rate is known from experiments \cite{tabeling}, \cite{carnevaleweiss}.

Another development of this analytical programme which can lead to
comparison with experiment and numerical simulation is connected with the
statistics of high amplitude, localized vortices, in a turbulent fluid
field. As mentioned above, high amplitude of vorticity concentration are
associated with large magnitude of $\rho $ (\ref{eqbun}). As explained
above, the mapping FT to CMC surfaces suggests $\left( \sqrt{\kappa }\rho
\right) ^{-1}=\kappa _{2}-\kappa _{1}$ which means that an approximative
localization of high-$\omega $ peaks can be obtained from the localization
of the umbilic points $\kappa _{1}=\kappa _{2}$. This can only be
approximative, \emph{i.e.} $\left\vert \kappa _{1}-\kappa _{2}\right\vert
<\Lambda $ since in fact umbilic points cannot exist on the torus (\emph{i.e.%
} the basic rectangle of the double periodic solutions of the sinh-Poisson
equation). One can take $\Lambda $ a fraction of $H$. Now, starting from a
CMC surface we consider a statistical ensemble of its realizations under
random perturbations. For Gaussian statistics of the perturbations of the
surface it is possible to determine the average density of the
randomly-occuring umbilic points \cite{umbilicberry}. This also provides the
average density of the localized, high magnitude concentration of vorticity
in a turbulent field.

\bigskip

\section{Summary}

The presence of convective structures in a turbulent field is recognized as
a frequent experimental regime in fluids and plasmas. It is also a difficult
theoretical problem. In this work we have examined a series of models which,
placed together, can lay the basis for a systematic analytical investigation
of this particular regime. The characteristic of a possible approach, as we
have tried to propose, is the need to work in the close proximity of exactly
integrable states, identified by variational methods (extremizing
functionals), not by conservation equations. It is determined, in this way,
a set of structures. They are known to have emerged in the late phase of
turbulence relaxation and for this reason, can be approximately taken as
existing even in the dynamical regime which precedes the final static state.
The way to do this analytically requires however a complicated apparatus.
The uniton solution appears to be a good direction of investigation.\ This
is part of a series of models that, beyond their apparent differences, are
strongly related: Euler fluid, point-like vortices, field theory of matter
and Chern-Simons gauge field, constant mean curvature surfaces. As we have
explained before, each model provides a certain realization of the regime
preceding the final state: (1) a set of fluid vortices moving slowly in a
weak turbulent field; (2) the matter and gauge fields of FT evolve in close
proximity of the self-duality state; (3) a set of unitons move on the
manifold of moduli space of solutions of the chiral model, with random
perturbations relative to the geodesic trajectory; (4) a constant mean
curvature surface perturbed by random deformations returns through local
curvature flows to the CMC state.

The chiral model, strongly related to constant mean curvature surfaces and
to the self-dual equations of the FT, offers the structures we need
(unitons) and their dynamic in time. Practical applications are possible and
in some cases the result can be confronted with observations.

\bigskip

{\bf Acknowledgments}
This work has been partly supported by grants WPJET1-RO-c and WPENR-RO-c of the Romanian Ministry of National Education and Scientific Research.

\bigskip

{\bf Bibliography}

\end{document}